\def\Granada{Instituto Carlos I de F\'\i sica Te\'orica y Computacional, 
Facultad de Ciencias, Universidad de Granada, Campus de Fuentenueva,
Granada 18002, Spain}

\def\Valencia{IFIC, Centro Mixto Universidad de Valencia-CSIC, Burjasot
              46100-Valencia,Spain.}

\def\INFN{Dipartimento di Scienze Fisiche and INFN, Sezione di Napoli,
      Mostra d'Oltremare, Pad. 19, 80125, Napoli, Italia.}

\def\nn{\nonumber}

\def\ni{\noindent}

\def\GC{{\cal G}_C}
\def\XL{\tilde{X}^L}
\def\XR{\tilde{X}^R}

\def\Gt{$\tilde{G}\,$}

\documentstyle[twoside,IJMPB]{article}
\textwidth=5truein
\textheight=7.8truein

\runninghead{V. Aldaya, J. Guerrero and G. Marmo}
{ Higher-order differential operators  on a Lie group
 $\ldots$}
\titleline{ HIGHER-ORDER DIFFERENTIAL OPERATORS \\ ON A LIE GROUP 
AND QUANTIZATION}
\vspace{0.37truein}
\writers{V. ALDAYA}
\place{\Granada\ and \\ \Valencia}
\vglue 10pt
\writers{J. GUERRERO}
\place{\Granada\ and \\ \INFN}
\vglue 10pt
\centerline{\eightrm and}
\writers{G. MARMO}
\place{\INFN}
\vspace{0.225truein}
\abstracts{\noindent This talk is devoted mainly to the concept of 
higher-order polarization on
a group, which is introduced in the framework of a Group Approach to 
Quantization,
as a powerful tool to guarantee the irreducibility of quantizations 
and/or representations of Lie groups in those  anomalous cases 
where the Kostant-Kirilov co-adjoint method or the Borel-Weyl-Bott
representation algorithm do not succeed. 
}{}{}

\setcounter{footnote}{0}





\section{Group Approach to Quantization}
We start with a Lie group $\tilde{G}$ which is a $U(1)$-bundle with 
base manifold $M$. A one-form $\tilde{\theta}\equiv \Theta$ on $\tilde{G}$ 
will be naturally selected (see below) among the components of the 
left-invariant, Lie Algebra valued, 
Maurer-Cartan 1-form, in such a way that $\Theta(\tilde{X}_0)=1$ and 
$L_{\tilde{X}_0}\Theta=0$, where $\tilde{X}_0$ is the vertical vector field.
$\Theta$ will play the role of a 
connection 1-form associated with the $U(1)$-bundle structure 
\cite{23,Ramirez}.

Let us consider firstly the case of a Lie group $\tilde{G}$ which is a 
central extension of a Lie group $G$ by $U(1)$, defining a principal bundle 
$\pi:\tilde{G}\rightarrow G$. Such a central extension 
is characterized by a Lie group 2-cocycle $\xi:G\times G\rightarrow R$ in 
terms of which the group law $\tilde{g}''=\tilde{g}'\tilde{*}\tilde{g}$ is 
written 
as $(g'',\zeta'')=(g'*g,\zeta'\zeta e^{i\xi(g',g)})$, or the corresponding
Lie algebra 2-cocycle 
$\Sigma: {\cal G}\times {\cal G}\rightarrow R$, where ${\cal G}$ is the 
Lie algebra of $G$. Thus, there is a unique, up to an exact 1-form $d\lambda$, 
left-invariant 1-form $\Theta$ on $\tilde{G}$ such that 
$d\Theta|_e=\tilde{\Sigma}$, where $\tilde{\Sigma}$ is the pull-back of 
$\Sigma$ by $\pi$.
The ambiguity in the definition of $\Theta$ only corresponds to the freedom
of adding a coboundary $\delta\lambda(g',g)=\lambda(g'*g)-\lambda(g')-
\lambda(g)$ to the cocycle. The fuction $\lambda$ proves to be a function 
with no linear terms in any canonical co-ordinate system at the identity.  

Now, we consider the case of a symmetry Lie group $G$ with trivial cohomology,
$H^2(G,U(1))=0$, but wearing a principal $H$-bundle structure ($H$=$U(1)$ 
or $R$) with projection $\pi:G\rightarrow M$, and let 
$\{\sigma_{\alpha\beta}\}$ be a 
\v{C}ech cocycle (which is unique save for a \v{C}ech coboundary) on a open 
covering $\{U_\alpha\}$ of $M$, defining this fibration. The composition
$\{\sigma_{\alpha\beta}\circ\pi\}$ turns out to be a Lie group cocycle (a 
coboundary in fact) \cite{Formal} defining a trivial central extension 
$\tilde{G}\equiv G\tilde{\otimes}U(1)$  whose Lie algebra structure constant
are, nevertheless, non-trivially modified, thus defining a connection 1-form 
$\Theta$ as in the previous case.
This type of not-so-trivial central extensions will be called 
pseudo-extensions 
and the corresponding coboundaries, pseudo-cocycles. These coboundaries 
constitute
a subgroup of the group of coboundaries, the pseudo-cohomology group, in 
correspondence (one-to-one at the Lie algebra level) with the true 
cohomology group of the 
contraction (in the sense of In\"on\"u and Wigner) of $G$ with respect 
to the subgroup $H$ \cite{Saletan,Pseudo}. 

In the general case, including infinite-dimensional semi-simple Lie groups, 
for which the Whitehead lemma does not apply, the principal bundle with 
connection $(\tilde{G},\Theta)$ should be constructed by applying both 
procedures to $G$, i.e. the group law for $\tilde{G}$ will contain cocycles
as well as pseudo-cocycles (see \cite{Virasoro,Formal,Anomalias}). 

We then start from a Lie group $\tilde{G}$ with connection 1-form $\Theta$ 
defined as before, verifying $i_{X_0}\Theta=1$, 
$X_0$ being the infinitesimal generator
of $U(1)$, or the fundamental vector field of the principal bundle.
Since $\Theta$ is left-invariant it will be preserved ($L_{X^R}\Theta=0$)
by all right-invariant vector fields (generating finite left 
translations) on $\tilde{G}$. These vector fields are candidates to be
infinitesimal generators of unitary transformations. However, to define the 
space of functions on which they should act, we first notice that by 
requiring

\begin{equation}
L_{X_0}\psi=i\psi
\end{equation}

\ni we select equivariant functions on $\tilde{G}$, which are associated 
with sections of the $U(1)$-bundle $\tilde{G}$ over $G$.
To make the action of the right-invariant vector fields on equivariant 
functions irreducible, any operator commuting with them must be 
trivialized, and this is achieved by  polarization conditions.

Two vector fields $\tilde{X}^L,\tilde{Y}^L$ (or the corresponding group 
parameters in local 
co-ordinates) are conjugated if 
$\tilde{\Sigma}(\tilde{X}^L,\tilde{Y}^L)\neq 0$. We will restrict 
ourselves to finite-dimensional Lie groups or infinite-dimensinal ones 
possessing a countable basis of generators for which, for arbitrary 
fixed $\tilde{X}^L$, $\tilde{\Sigma}(\tilde{X}^L,\tilde{Y}^L)=0$ except for a 
finite number of vectors 
$\tilde{Y}^L$ (finitely non-zero cocycle). In these cases the skewsymmetric 
form $\tilde{\Sigma}$, seen as a pairing 
$\tilde{\Sigma}:\tilde{\cal G}\times\tilde{\cal G}\rightarrow R$, can always 
be taken to normal form defining  canonically conjugate pairs of 
vector fields as well as those without symplectic character. For instance, 
for the Galilei group the non-symplectic generators would be those associated 
with time, 
spatial rotations or the vertical generator itself. The intersection of
$Ker\tilde{\Sigma}\equiv Ker d\Theta$ and $Ker\Theta$ (horizontality) in 
${\cal X}^L(\tilde{G})$ 
defines a vector subspace, which proves to be a subalgebra, the 
{\bf characteristic subalgebra} ${\cal G}_C$ generating the characteristic 
module of $\Theta$. 

A {\bf first-order polarization} or just {\bf polarization} ${\cal P}$ is 
defined as a maximal horizontal 
left subalgebra. The horizontality condition means only 
that the $U(1)$ generator is excluded from the polarization, a 
fact which will be relevant for further generalizations. 

A polarization may have non-trivial intersection with the characteristic 
subalgebra. 
We say that a polarization is {\bf full (or regular)} if it contains the 
whole characteristic subalgebra. We also say that a polarization is 
{\bf symplectic} if it contains ``half" of the symplectic vector fields, i.e. 
one of each pair of canonically conjugate vector fields.  

Among all complex-valued functions on $\tilde{G}$ we select the 
{\bf wave functions} as those which
are constant along maximal isotropic integral surfaces of the polarization, 
i.e. $X^L\Psi=0, \forall X^L\in{\cal P}$, and satisfy the equivariance
condition $L_{X_0}\psi=i\psi$. 

\vskip 0.4cm

\ni {\large \it The Harmonic Oscillator}

\vskip 0.3cm 

For a simple example, let us consider the case of the harmonic oscillator.
The quantization group \cite{23} is a symmetry group 
$\tilde{G}_{(m,\omega)}$ which goes to the centrally extended
Galilei group as $\omega\rightarrow 0$. To simplify the posterior 
search for a polarization, we write the group law in (otherwise standard)
coordinates $c\equiv\sqrt{\frac{m}{2\omega}}(\omega x+iv), c^*\equiv
\sqrt{\frac{m}{2\omega}}(\omega x-iv)$, and $\eta\equiv e^{i\omega t}$.
The group law is:
\begin{eqnarray}
c''&=&c'\eta^{-1}+c  \nn \\
{c^*}''&=&{c^*}'\eta+c^*  \nn \\
\eta''&=&\eta'\eta \\
\zeta''&=&\zeta'\zeta e^{\frac{i}{2}[ic'c^*\eta^{-1}-i{c^*}'c\eta]} \nn
\end{eqnarray}

\ni For the left- and right-invariant vector fields we find 
($X_0\equiv\Xi$):

\begin{equation}\begin{array}{lcl}{\XL}_{\eta}&=&i\eta\frac{\partial \,}
{\partial \eta}-ic\frac{\partial \,}{\partial c}+
ic^*\frac{\partial \,}{\partial c^*} \\
 {\XL}_c&=&\frac{\partial \,}{\partial c}-\frac{i}{2}c^*\Xi \\ {\XL}_{c^*}&=&
\frac{\partial \,}{\partial c^*}+\frac{i}{2}c\Xi \\ {\XL}_{\zeta}&=&
i\zeta\frac{\partial \,}{\partial \zeta}\equiv \Xi
  \end{array} \;\;\;\; \begin{array}{lcl}{\XR}_{\eta}&=&i\eta\frac{\partial 
\,}{\partial \eta} \\
{\XR}_c&=&\eta^{-1}\frac{\partial \,}{\partial c}+\frac{i}{2}\eta^{-1}c^*\Xi 
\\ {\XR}_{c^*}&=&\eta\frac{\partial \,}{\partial c^*}-
\frac{i}{2}\eta c\Xi \\ {\XR}_{\zeta}&=&i\zeta\frac{\partial \,}
{\partial \zeta}\equiv \Xi
\end{array}
\end{equation}
\ni with commutation relations
\begin{eqnarray}
\left[\XL_c,{\XL}_{c^*}\right]&=&i\Xi \nn \\
\left[\XL_{\eta},\XL_c\right]&=&i\XL_c \\
\left[\XL_{\eta},\XL_{c^*}\right]&=&-i\XL_{c^*} \nn
\end{eqnarray}
\ni The  quantization form, dual to $\XL_{\zeta}$ is
\begin{eqnarray}
\Theta=\frac{i}{2}[c^*dc&-&cdc^*]-cc^*\frac{d\eta}{i\eta}+
\frac{d\zeta}{i\zeta}
 \\ (d\Theta=d\Theta_{PC}&=&idc^*\wedge dc-\omega dH\wedge dt) \nn
\end{eqnarray}

\ni The characteristic subalgebra is generated by the time generator:
$\GC=<\XL_{\eta}>$ and a (full and symplectic) polarization is 
${\cal P}=<\XL_{\eta},\XL_c>$. The wave functions are then solutions of 
the following equations:

\begin{eqnarray}
\Xi.\Psi&=&i\Psi\Rightarrow\Psi(\zeta,\eta,c,c^*)=\zeta \Phi(\eta,c,c^*) 
\nn \\
{\XL}_c.\Psi=0&\Rightarrow&\Phi=e^{-\frac{cc^*}{2}}\varphi(c^*,\eta)  \nn \\
{\XL}_{\eta}.\Psi=0&\Rightarrow&i\frac{\partial\varphi}{\partial t}=
\omega c^*\frac{\partial \varphi}{\partial c^*} \nn
\end{eqnarray}
\ni with the general expression

\begin{equation}
 \Phi=e^{-\frac{cc^*}{2}}\sum_n A_n{c^*}^n e^{-in\omega t}\nn  
\end{equation}

\ni In particular, coherent states turn out to be

\begin{equation}
|c>\equiv\sum_{n=0}^{\infty}\Phi_n(c,c^*)^*|n> \,\,\,,\,
\Phi_n=e^{-\frac{|c|^2}{2}}\frac{{c^*}^n}{\sqrt{n!}}\,, \,\,\, |n>\equiv
     \frac{(\hat{c}^{\dag})^n|0>}{\sqrt{n!}} \,,
\end{equation}

\ni where $\hat{c}^{\dag}\equiv - \XR_{c}$. Notice that the vacuum $|0>\equiv 
e^{-\frac{cc^*}{2}}$ is characterized 
by being nullified by $\hat{E}\equiv i\omega\XR_{\eta}$ and 
$\hat{c}\equiv\XR_{c^*}$. 

\section{Algebraic Anomalies}

In Sec.2, we introduced  the concept of full and symplectic polarization 
subalgebra intended to reduce the Bohr representation. 
It contains half of the symplectic variables as well as the entire 
characteristic subalgebra. If the full reduction is achieved, 
the whole set of physical 
operators can be rewritten in terms of the basic ones, i.e. those 
associated with the symplectic variables or, in other words,
the operators outside the characteristic subalgebra. For instance, the 
energy operator for the free particle can be written as 
$\frac{\hat{p}^2}{2m}$, the angular momentum in 3+1 dimensions is the 
vector product ${\bf \hat{x}}\times {\bf \hat{p}}$, or the energy for 
the harmonic oscillator is $\hat{C}^{\dag}\hat{C}$.

However, the existence of a full and symplectic polarization is not guaranteed.
We define  an {\bf anomalous} group \cite{Anomalias} as a a central extension 
\Gt which do not admit a full and symplectic polarization for certain values 
of the (pseudo-)cohomology parameters, called the {\bf classical} values of 
the anomaly.
Anomalous groups feature another set of values of the (pseudo-)cohomology 
parameters, called the {\bf quantum} values of the anomaly, for which the 
carrier space associated
with a full and symplectic polarization contains an invariant subspace.
For the classical values of the anomaly, the classical solution manifold 
undergoes a reduction in dimension thus increasing the number of 
(non-linear) relationships among Noether invariants,  whereas for the quantum 
values the number of 
basic operators decreases on the invariant (reduced) subspace due to
the appearance of (higher-order) relations among the quantum operators.

We must remark that the anomalies we are dealing with in this paper are
of {\it algebraic} character in the sense that they appear at the Lie algebra 
level,
and must be distinguished from the {\it topologic anomalies} which are
associated with the non-trivial homotopy of the (reduced) phase space 
\cite{Frachall}.

The non-existence of a full and/or symplectic polarization is traced back to the
presence in the characteristic subalgebra associated with certain values of 
the (pseudo-)cohomology (the classical values of the anomaly) of some 
element the adjoint action of which is not diagonalizable in the ``$x-p$" 
algebra subspace. The anomaly problem here presented parallels that of the
non-existence of invariant polarizations in the Kirillov-Kostant co-adjoint
orbits method \cite{Gotay}, and the conventional anomaly problem in
Quantum Field Theory which manifests itself through the appearance of
central charges in the quantum current algebra, absent from the classical
(Poisson bracket) algebra \cite{Jackiw}.
 
The full reduction of  representations in the anomalous case will be achieved 
by means of a generalized concept (higher-order) polarization (see below). 

\vskip 0.3cm

\ni {\large \it The Schr\"odinger group}

\vskip 0.2cm

To illustrate the Lie algebra structure of an anomalous symmetry, let us 
consider the example of the Schr\"odinger group.
It is the symmetry of the Schr\"odinger equation for generic 
potential $Ax^2+Bx+C$ and its Lie algebra (a central extension indeed) has a 
very simple realization as 
the Poisson subalgebra \cite{Niederer}
\begin{eqnarray}
&&\{1,x,p,p^2,x^2,x\cdot p\} \;\;\;\; \nn \\
 &or& \\
&&\{1,x,p,p^2+x^2,x^2,x\cdot p\} \;\;\; \nn
\end{eqnarray}

\ni The two trivially equivalent versions of the same algebra reflect the 
existence of two non-equivalent classes of representations, the first one 
supported on the
wave functions of the free particle, the second on the wave functions of the
harmonic oscillator. 

Let us consider the harmonic oscillator-like version of the Schr\"odinger
algebra and define, accordingly, coordinates $c,c^*$ in the standard way:
$x\equiv\frac{1}{\sqrt{2m\omega}}(c+c^*), p\equiv\frac{-im\omega}
{\sqrt{2m\omega}}(c-c^*)$,
in terms of which the classical Poisson-algebra generators are written:

\begin{equation}
\{1,c,c^*,cc^*,c^2,{c^*}^2\}\equiv\{1,c,c^*,\eta,z,z^*\}
\end{equation}

\ni where we have denoted the quadratic generators by linear, independent 
variables
since all the generators in an abstract algebra are independent. We can think 
of the 
Schr\"odinger algebra as that of the harmonic oscillator where the $sl(2,R)$
algebra substitutes the $u(1)$ algebra associated with time.

Let us write the Poisson brackets to analyse the anomalous
structure: 

\begin{equation}\begin{array}{rcr}\{\eta,c\}&=&-c \\
 \{\eta,c^*\}&=&c^*\\ \{\eta,z\}&=&-2z \\ \{\eta,z^*\}&=&2z^* \\ \{c,z^*\}&=&
   \frac{1}{\sqrt{2}}c^* 
  \end{array} \;\;\;\; \begin{array}{rcr}\{c^*,c\}&=&1 \\
\{z^*,z\}&=&-\frac{1}{2}\eta \\ \{c,z\}&=&0 \\ \{c^*,z^*\}&=&0 \\ \{c^*,z\}&=
&\frac{1}{\sqrt{2}}c
\end{array}
\end{equation}

\ni The last line above prevents the existence of a full and symplectic 
polarization; we can find only a symplectic (non-full) polarization (and the 
conjugate one) which in terms 
of the corresponding left-invariant vector fields \cite{Anomalias}  becomes

\begin{equation}
{\cal P}=<\XL_c,\XL_{\eta},\XL_z>\,, \label{non-full}
\end{equation}

\ni and a full (non-symplectic) polarization 

\begin{equation}
{\cal P}_C=<\XL_{\eta},\XL_z,\XL_{z^*}>\,, \label{non-symplectic}
\end{equation}
 
\ni Quantizing with the non-full polarization (\ref{non-full}) results in
a breakdown of the naively expected correspondence between the operators 
$\hat{z},\hat{z}^{\dag}$ and the basic ones:
\begin{eqnarray}
\hat{z}&\not\sim&\hat{c}^2  \nn \\
\hat{z}^{\dag}&\not\sim&\hat{c}^{\dag}{}^2 
\end{eqnarray}

Unlike in the classical case, the operators 
$(\hat{\eta}),\hat{z},\hat{z}^{\dag}$
behave independently of $\hat{c},\hat{c}^{\dag}$.  
The operators $\hat{z},\hat{z}^{\dag}$ seem to have symplectic content as if 
they were canonically-conjugate (basic) operators.

The quantization with the non-symplectic polarization (\ref{non-symplectic})
leads to an unconventional representation in which the wave functions 
depend on both $c$ and $c^*$, but it is nevertheless irreducible. The 
operators  $\hat{z},\hat{z}^{\dag}$, neither, are expressed in terms of
$\hat{c},\hat{c}^{\dag}$. 

\vskip 0.3cm

\ni {\large \it The Virasoro group}

\vskip 0.2cm

Let us comment very briefly on the relevant, although less intuitive,
example of the infinite-dimensional Virasoro group. Its Lie algebra can 
be written as

\begin{equation}
\left[\XL_{l_n},\XL_{l_m}\right]=-i(n-m)\XL_{l_{n+m}}-\frac{i}{12}(cn^3-c'n)
\Xi\,, 
\end{equation}

\ni where $c$ parametrizes the central extensions and $c'$ the 
pseudo-extensions. As is well known, for the particular case in which
$\frac{c'}{c}=r^2\,,r\in N, r>1$, the co-adjoint orbits are not K\"ahlerian.
In the present approach, this case shows up as an algebraic anomaly. In 
fact, the characteristic subalgebra is given by 
$\GC=<\XL_{l_0},\XL_{l_{-r}},\XL_{l_{+r}}>$, which is not fully 
contained in the non-full (but symplectic) polarization 
${\cal P}^{(r)}=<\XL_{l_{n\leq 0}}>$. 
A detailed description of the representations of the Virasoro group can be 
found in \cite{Virasoro} 
and references therein.

\section{Higher-order Polarizations}

In general, to tackle  situations like those mentioned above, it is necessary 
to generalize the notion of polarization. Let us consider the universal 
enveloping 
algebra of left-invariant vector fields, 
${\cal U}\tilde{\cal G}^L$. We say that a subalgebra ${\cal A}$ of 
${\cal U}\tilde{\cal G}^L$ is 
{\bf horizontal} if it does not contain the vertical generator $X_0$.

Then we define a {\bf higher-order polarization} ${\cal P}^{HP}$ as a maximal 
horizontal 
subalgebra of ${\cal U}\tilde{\cal G}^L$. With this definition a higher-order 
polarization contains the maximum number of conditions compatible 
with the equivariance condition of the wave functions and with the action 
of the physical operators (right-invariant vector fields).

We notice that now the vector space of functions annihilated by a
maximal higher-order polarization is not, in general, a ring of 
functions and therefore there is no corresponding foliation; that is, 
they cannot be characterized by saying that they are constant along 
submanifolds. If this were the case, it would mean that the 
higher-order polarization was the enveloping algebra of a first-order 
polarization and, accordingly, we could consider the submanifolds 
associated with this polarization. In this sense the concept of 
higher-order polarization generalizes and may replace that of first-order 
polarization.

We arrive at the formulation of our main general theorem, which was firstly
proven for the particular case of the Virasoro group in Ref. \cite{Virasoro}
and more generally in Ref. \cite{Marmo}

\ni {\bf  Theorem}: {\it Let ${\cal P}^{HO}$ be a higher-order polarization
on \Gt. On subspaces characterized by}

\begin{equation}
L_{X_0}\psi=i\psi\,,\;\;A.\psi=0\;\; \forall  A\in {\cal P} 
(polarization) \label{theorem}
\end{equation}

\ni {\it all the right-invariant vector fields} $\XR$ 
{\it act irreducibly. Therefore the present quantization procedure 
gives rise to an irreducible representation of the group} 
$\tilde{G}$, {\it provided it is connected and simply connected}.

The definition of higher-order polarization given above is quite general. In 
all studied examples higher-order polarizations adopt a more definite 
structure 
closely related to given first-order (non-full and/or non-symplectic) ones.
According to the until now studied cases higher-order polarizations can be 
given a more operative definition: {\it A higher-order polarization is a 
maximal horizontal subalgebra of ${\cal U}\tilde{\cal G}^L$ the vector 
field content of which is a first order polarization}. 

To see how a higher-order polarization operates in practice, we come back
to the case of the Schr\"odinger group and the representation associated
with the non-full polarization (\ref{non-full}) for which the operators
$\hat{z},\hat{z}^{\dag}$ are basic. However, for a particular value (the 
quantum 
value of the anomaly) $k=\frac{1}{4}$, the representation of the Schr\"odinger
group {\it becomes reducible}, although not completely reducible and, 
{\it on the invariant subspace}, the operators $\hat{z},\hat{z}^{\dag}$ do 
really 
express as  $\hat{c}^2,\hat{c}^{\dag}{}^2$ except for numerical 
proportionality 
constants. The invariant subspace is constituted by the solutions
of a second-order polarization which exists only for $k=\frac{1}{4}$:

\begin{equation}
{\cal P}^{HO}=<\XL_c,\XL_{\eta},\XL_z, \XL_{z^*}-\alpha(\XL_{c^*})^2>
\end{equation}

\ni where the constant $\alpha$ is forced to acquire the value $\alpha=
\frac{1}{2\sqrt{2}}$. Physical applications of this particular representation 
are
found in Quantum Optics \cite{Optica} although no reference to the connection
between anomalies and the restriction of $k$ has been made.

In a similar way, in the case of the Virasoro group, for particular values
of the parameters $c,c'$ or equivalently $c,h\equiv \frac{c-c'}{24}$ given
by the Kac formula \cite{Kac}, the ``quantum values" of the anomaly, 
the representations given by the first order non-full (symplectic) 
polarizations are reducible since there exist invariant subspaces 
characterized by certain higher-order polarization equations 
\cite{Virasoro}. Note that there is no one-to-one correspondence 
between the values of $c'/c$ characterizing the coadjoint orbits of the 
Virasoro group (the classical values of the anomaly) and the values 
allowed by the Kac formula (the quantum values of the anomaly), a fact 
which must be interpreted as a breakdown of the notion of classical 
limit.

Let us remark, finally, that higher-order polarizations can also be applied 
to non-anomalous groups to obtain a different, although equivalent, 
realization of an irreducible representation associated with a first-order
full and symplectic polarization. This is the case of the configuration-space
realization which cannot be obtained by means of first-order polarizations.
Thus, for instance, for the free non-relativistic particle and for the
harmonic oscillator, the higher-order polarization 
${\cal P}^{HO}=<\XL_p, \XL_t-\frac{i\hbar}{2m}(\XL_x)^2>$ leads to the 
corresponding 
quantizations in configuration space \cite{Anomalias}. In the relativistic case
infinite-order polarizations are required \cite{Position,Relativistic}. The 
{\it modus operandi} in these cases, according to the operative definition 
of higher-order polarization, is to start with a non-full symplectic 
polarization containing the generators associated with the variables 
canonically conjugated to $x$ (or the ones associated with any other 
realization), ${\cal P}=<\XL_p>$ in this case, and complete it with elements 
of 
the left enveloping algebra to substitute each one of the generators of the 
characteristic algebra lacking in the original first-order 
polarization, in such a way that all of them close a horizontal subalgebra; 
$\XL_t-\frac{i\hbar}{2m}(\XL_x)^2$ substitutes to $\XL_t$ 
in this case. In a more general (relativistic) system $\XL_t$ would be
substituted by an infinite series in the left enveloping 
algebra starting by the first-order term $\XL_t$ \cite{Relativistic}.   

\nonumsection{Acknowledgment}
This work was partially supported by the Direcci\'on General de 
Investigaci\'on Cient\'\i fica y T\'ecnica.

\nonumsection{References}

\end{document}